\documentclass[aps]{revtex4}

\usepackage{graphicx}

\begin{document}

\title{Python for Education: Computational Methods for Nonlinear Systems}
\author{Christopher~R.~Myers$^{1}$ and James.~P.~Sethna$^{2}$}
\affiliation{$^{1}$Cornell Theory Center, Cornell University, Ithaca, NY 14853, USA\\
$^{2}$Laboratory of Atomic and Solid State Physics, Cornell University, Ithaca, NY 14853, USA}

\date{\today}

\begin{abstract}
We describe a novel, interdisciplinary, computational methods course that 
uses Python and associated numerical and visualization libraries to enable 
students to implement simulations for a number of different course modules.
Problems in complex networks, biomechanics, pattern formation, and gene 
regulation are highlighted to illustrate the breadth and flexibility 
of Python-powered computational environments.
\end{abstract}

\maketitle

\section*{Introduction}
\label{Sec-Introduction}

Computational science and engineering (CSE) involves the integration
of a number of different techniques, requiring expertise in data
structures, algorithms, numerical analysis, programming methodologies,
simulation, visualization, data analysis, and performance
optimization.  The CSE community has embraced Python as a platform for
attacking a wide variety of research problems, in part because of
Python's support for easily gluing together tools from different
domains to solve complex problems.  Teaching the theory and practice
of CSE requires touching on all the subjects mentioned above, in the
context of important and interesting scientific problems.  Many of the
same advantages that Python brings to CSE research make it useful for
teaching too.  Traditionally, courses have tended to focus more
narrowly on particular aspects of CSE, such as numerical analysis,
algorithms, or high-performance computing.  In developing a new,
broadly focused laboratory course in computational science,
engineering and biology, we have sought to introduce students to the
wide swath of techniques necessary to do effective research in CSE.
Python and its many batteries serve remarkably well in this endeavor.

\emph{Computational methods for nonlinear systems} is a graduate
computational science laboratory course jointly developed and taught
by us.  We initiated course development in the summer of 2004 to
support the curricular needs of the Cornell IGERT program in nonlinear
systems, a broad and interdisciplinary graduate fellowship program
aimed at introducing theoretical and computational techniques
developed in the study of nonlinear and complex systems to a range of
fields.  The focal themes of the IGERT program span a number of areas
- including complex networks, biological locomotion and manipulation,
pattern formation, and gene regulation - broadly interpreted in the
context of complex systems and nonlinear dynamics.  These themes form
the core of our course curriculum, augmented with other problems of
interest arising in the fields of statistical mechanics, applied
mathematics, and computer science.

The format of the course is somewhat unusual.  As a computational
laboratory course, it provides relatively little in the way of
lectures: we prefer to have students learn by doing, rather than
having us tell them how to do things.  The course is autonomous,
modular, and self-paced: students choose computational modules to work
on from a large (and hopefully growing) suite of those available, and
then proceed to implement relevant simulations and analyses as laid
out in the exercise.  We provide \emph{Hints} files to help the
students along: these consist of documented skeletal code that the
students are meant to flesh out.  (In practice, we develop a module
ourselves, document each of the relevant pieces using Python's
docstrings, and then replace all the code bodies with the Python
keyword \verb+pass+ so that the students can repopulate those code
bodies themselves.)  We have written several different visualization tools
to provide visual feedback.  We find these help to 
engage the students in new problems and are useful in code debugging.

Python is a useful language for teaching for several reasons (even 
though most of our incoming students have had no previous
experience with Python).  Its clean syntax enables students to learn
the language quickly, and allows us to provide concise 
programming hints in our documented code fragments.  Python's dynamic
typing and high-level, built-in datatypes enables students to get
programs working quickly, without having to struggle with type
declarations and compile-link-run loops.  Since Python is interpreted,
students can learn the language by executing and analyzing individual
commands, and we can help them debug their programs by working with
them in the interpreter.

One of the other key advantages that Python brings to scientific
computing is the availability of many packages supporting numerical
algorithms and visualization.  While some of our exercises require
development of algorithms from scratch, others rely on established
numerical routines implemented in third-party libraries.  It is of
course important to understand the fundamentals of algorithms, error
analysis, and algorithmic complexity, but it is also useful to know
when and how to use existing solutions that have been developed by
others.  We make heavy use of the {numpy}\cite{Numpy} and
{scipy}\cite{Scipy} packages, for construction of efficient arrays and
for access to routines for generation of random numbers, integration
of ordinary differential equations, root-finding, computation of
eigenvalues, etc.  We use {matplotlib}\cite{Matplotlib} for x-y
plotting and histograms.  We have written several visualization
modules that we provide to students, based on the {Python Imaging
Library (PIL)}\cite{PIL},  using PIL's ImageDraw
module to place graphics primitives within
an image, and the ImageTk module to paste an image into a Tk
window for real-time animation.  We recommend the use of the {ipython}
interpreter, which facilitates exploration by students\cite{IPython}.
And we have used {VPython}\cite{VPython} to generate three-dimensional
animations to accompany some of our modules.

\section*{Course modules}
\label{Sec_Course_modules}

There are too many course modules to describe in detail here, and we
refer interested readers to our course website \cite{CM4NS} for
information on all the modules, as well as access to problems, hints,
and answers.  (Many of the exercises have also been incorporated into
a new textbook written by one of us.\cite{Sethna2006})  Here, we
highlight a few of the modules, in order to illustrate both the
breadth of science that can be usefully taught with Python and variety
of tools and techniques that Python can bring to bear on such
problems.

\subsection*{Small world networks}

The study of complex networks has flourished over the last several
years as researchers have discovered commonalities among networked
structures that arise in diverse fields such as biology, ecology,
sociology, and computer science\cite{Barabasi2002}.  An interesting
property found in many complex networks is exemplified in the popular
notion of ``six degrees of separation'', which suggests that any two
people on earth are connected through at most roughly five
intermediate acquaintances.  Duncan Watts and Steve
Strogatz\cite{Watts1998} developed a simple model of random networks
that demonstrate this ``small-world'' property.  Our course module
enables students to construct small-world networks and to examine how
the average path length connecting two nodes decreases rapidly as
random, long-range bonds are introduced into a network consisting
initially of only short-ranged bonds (Figure 1).

Computationally, this module introduces students to data structures
for the representation of undirected graphs, object-oriented
encapsulation of those data structures, and graph traversal
algorithms.  Python makes the development of an undirected graph data
structure exceedingly simple, a point made long ago by Python creator
Guido van Rossum in one of his early essays on
Python\cite{VanRossum1998}.  In an undirected graph, nodes are
connected to other nodes by edges.  A simple way to implement this is
to combine the two cornerstones of container-based programming in
Python: lists and dictionaries.  In our \verb+UndirectedGraph+ class,
a dictionary of network neighbor connections (a neighbor dictionary)
maps a node identifier to a list of other nodes to which the reference
node is connected.  Because the graph edges are undirected, we
duplicate the connection information for each node: if an edge is
added connecting node 1 and node 2, the neighbor dictionary must be
updated so that node 2 is added to node 1's list of neighbors, and
vice versa.

\begin{figure}
\includegraphics[height=2in]{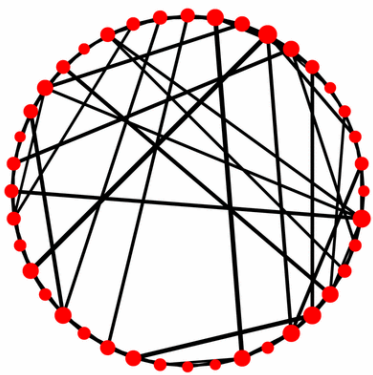}
\caption{\label{SmallWorldFig}
Node and edge betweenness in a model of small-world networks.  Nodes
(red dots) are connected by undirected edges (black lines).
Betweenness measures how central each node and edge is to the shortest
network paths connecting any two nodes.  In this plot, node diameter
and edge thickness are propotional to node and edge betweenness,
respectively.  (Our simple graph visualization tool uses the Python Imaging
Library.)
}
\end{figure}

We can of course hide the details of adding edges 
inside an \verb+AddEdge+ method defined on an \verb+UndirectedGraph+ class:

\begin{verbatim}
class UndirectedGraph:
    # ...
    def AddEdge(self, node1, node2):
        self.AddNode(node1)
        self.AddNode(node2)
        if node2 not in self.neighbor_dict[node1]:
            self.neighbor_dict[node1].append(node2)
        if node1 not in self.neighbor_dict[node2]:
            self.neighbor_dict[node2].append(node1)
\end{verbatim}

In the small-world networks exercise, we choose to label nodes simply
by integers, but Python's dynamic typing does not require this.  If we
were playing the ``Kevin Bacon game'' of searching for
shortest paths in actor collaboration networks, we could use our code
above to build a graph connecting names of actors (encoded as
strings).  This dynamic typing allows for significant code reuse (as
described below in the section on Percolation).  And it is worth
mentioning that, while our \verb+UndirectedGraph+ class is exceedingly simple
and built to support only the analyses relevant to our course module,
the same basic principles are at work in a much more comprehensive,
Python-based, graph construction and analysis package - named NetworkX
- that has been developed at Los Alamos National Labs.\cite{NetworkX}

\subsection*{Percolation}

Percolation is the study of how objects become connected (or
disconnected) as they are randomly wired together (or cut apart).
Percolation is an important and classic problem in the study of phase
transitions, and has practical relevance as well: considerable
interest over the years in percolation phenomena has come from the oil
and gas industry, for example, where one is interested in extracting a
fluid through a network of pores in rock.

Although percolation is traditionally studied on regular lattices, it
is a problem more generally applicable to arbitrary networks, and in
fact, we are able to reuse some of the code developed in the
small-world networks module to support the study of percolation.  As
noted above, Python's dynamic typing makes our definition of a node in
a graph very flexible; in a percolation problem on a lattice, we can
reuse our \verb+UndirectedGraph+ class described previously by making node
identifiers be lattice index tuples $(i,j)$.  We can thus easily make an
instance of bond percolation on a 2D square lattice of size $L$ (with
periodic boundary conditions) and bond fraction $p$:

\begin{verbatim}
def MakeSquareBondPercolation(L, p):
    g = UndirectedGraph()
    for i in range(L):
        for j in range(L):
            g.AddNode((i,j))
            if random.random() < p:
                g.AddEdge((i,j), ((i+1)%L, j))
            if random.random() < p:
                g.AddEdge((i,j), (i, (j+1)%L))
    return g
\end{verbatim}

\begin{figure}
\includegraphics[height=2in]{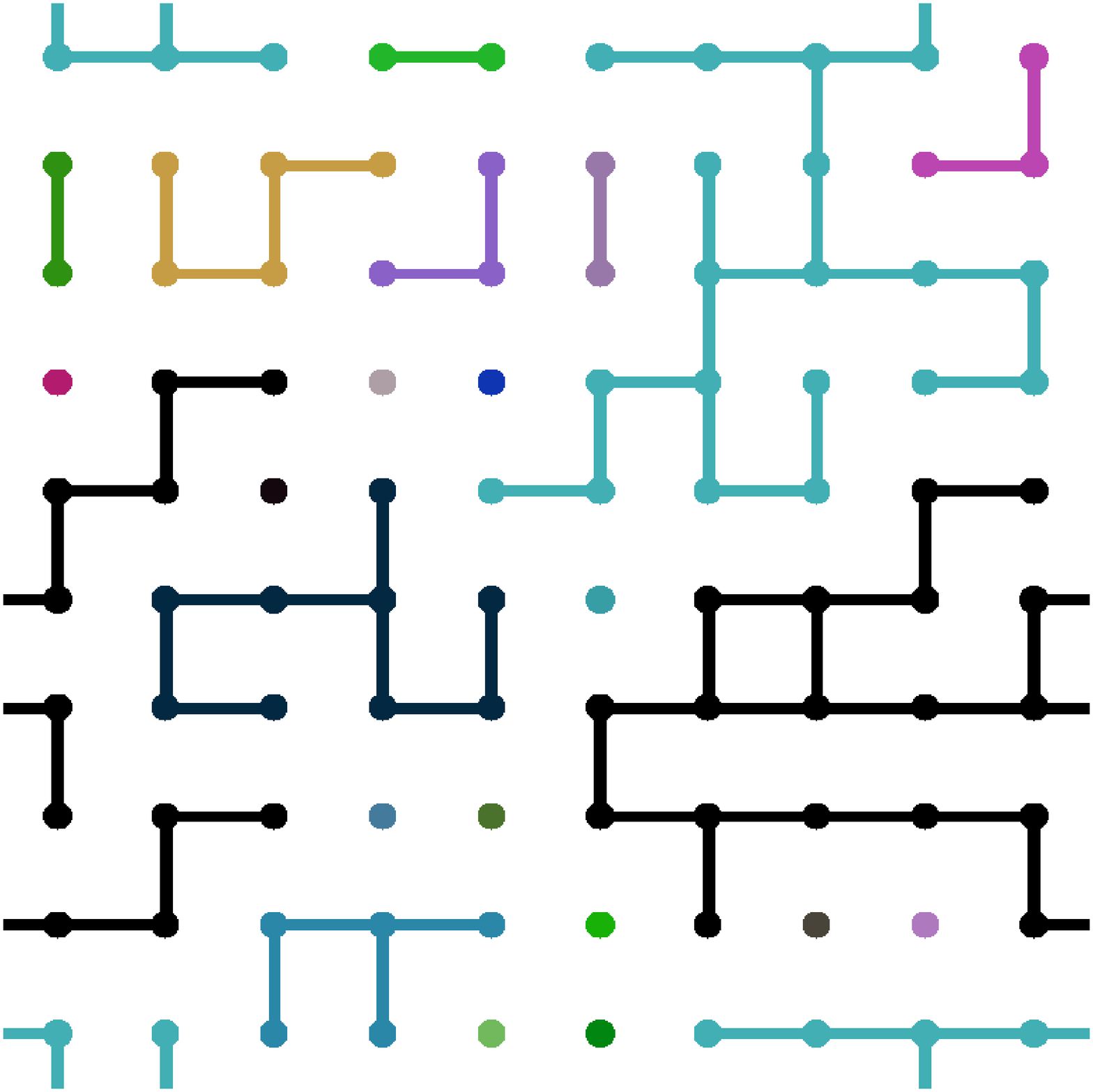}
\includegraphics[height=2in]{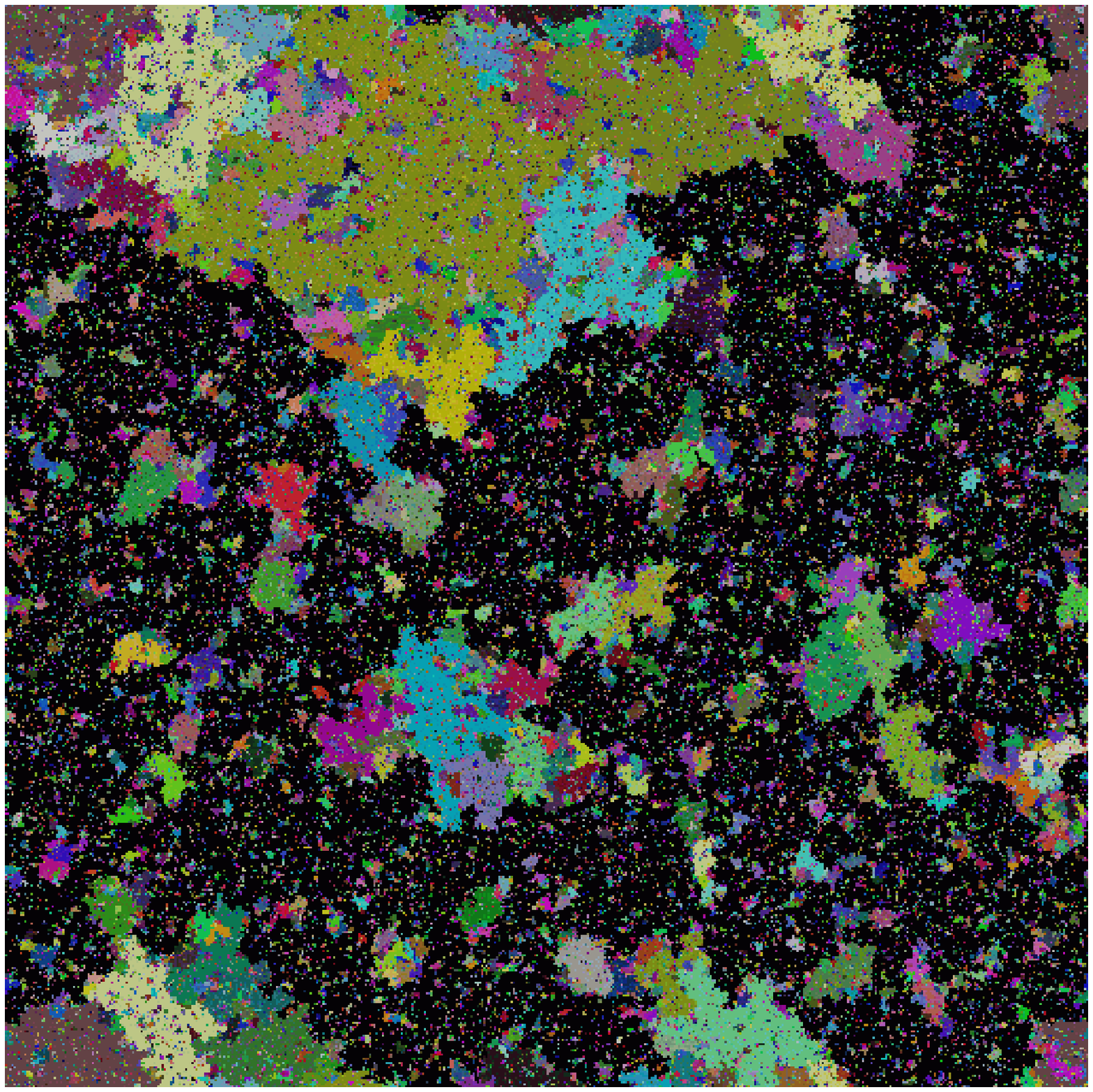}
\includegraphics[height=2in]{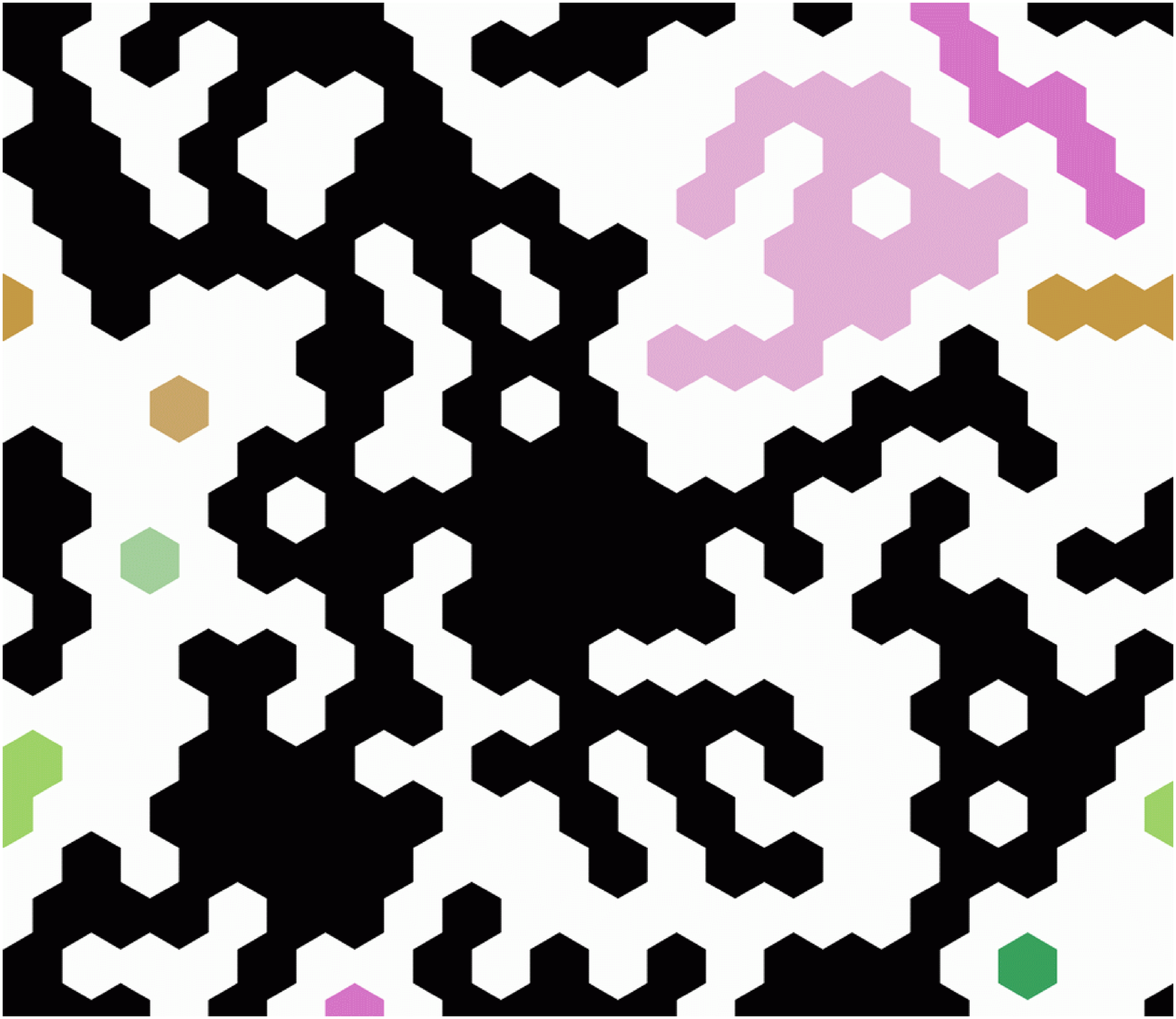}
\caption{\label{PercolationFig}
Two instances of bond percolation on a 2D square lattice, 
and an instance of site percolation on a hexagonal lattice.
In bond percolation, neighboring lattice points are connected with
probability $p$, and connected clusters in the resulting network are
identified via breadth-first search.  Separately clusters are colored
distinctly, for a 10x10 grid (left) and a 1024x1024 grid (middle).  In
site percolation (right), lattice sites are filled with probability
$p$, and clusters connect neighboring sites that are filled.  We study
both bond and site percolation to to introduce the concept of
universality of phase transitions.}
\end{figure}

Instances of percolation networks generated by this procedure are
illustrated in Figure 2.  Students use breadth-first search to identify
all connected clusters in such a network, and our PIL-based
visualization tool colors each separate cluster distinctly, taking as
input a list of all nodes in each cluster.  

The concept of universality of phase transitions is also introduced:
despite their microscopic differences, site-percolation on a 2D
hexagonal lattice and bond-percolation on a 2D square lattice are
indistinguishable from each other on long length scales, exhibiting
the same critical behavior (e.g., scaling exponents).  Scaling
collapses are a useful construct for revealing the universality of
phase transitions, and typically involve transforming the $x$ and $y$
axes in specified ways to get disparate data sets to ``collapse'' onto
one universal scaling form.  With Python, we can support 
such scaling collapses very flexibly by using the built-in
\verb+eval()+ function that evaluates expressions encoded as strings.
Rather than hard-coding particular functional forms for scaling collapses,
arbitrary mathematical expressions can be simply encoded and evaluated.

\subsection*{Biomechanics: The Walker}

Research in biomechanics aims to understand how living beings move,
and robotics and prosthetics are two important technological areas
that can benefits from advances in the field.  While much research in
robotics is focused on active sensing and control, Andy Ruina and
collaborators have been interested in passive biolocomotive systems,
which are more properly understood as dynamical systems than as
control systems.  The ``simplest walking model'' of Garcia et
al.\cite{Garcia1998} provides the basis of our Walker module.  This
model consists of a pair of legs connected at the hip (a double
pendulum), walking down an inclined ramp under the influence of
gravity, with a heelstrike that imparts angular momemtum to the Walker
as the swing leg strikes the floor and becomes the stance leg.  As a
warmup, students integrate the equation of motion for a single
pendulum under gravity, and compute the period of the motion as a
function of the initial pendulum angle.

The Pendulum and Walker modules introduce several important scientific
and computational aspects.  Ordinary differential equations (ODEs) 
describing the time evolution of the Pendulum and Walker 
need to be integrated forward in time.  In 
the context of the simpler Pendulum, we highlight the properties of 
accuracy, fidelity, and stability in numerical integration, having 
students explore errors introduced by a finite time step $\Delta t$.
We also highlight the need for event detection in many numerical 
integration problems.  In the Walker, for example, 
a heelstrike occurs when the swing leg hits the floor.
Accurately solving for the heelstrike collision involves transforming
to a new set of integration variables, where an appropriate
combination of the pendulum angles becomes the independent variable,
and time a dependent variable.  We then integrate backwards in angle
to find the time at which the heelstrike occurred.  We use the
scipy.integrate.odeint function to execute these integrations,
providing a function \verb+dydt+ that evaluates the instantaneous time
derivative of the Walker state vector field $\vec y$ and a function
\verb+dzdc+ that evaluates the instantaneous time derivative of the
transformed system for heelstrike detection (where the independent variable
is the ``collision variable'' $c = \phi-2\theta$).

\begin{verbatim}
def dydt(self, y,t):
    theta,thetaDot,phi,phiDot = y
    thetaDotdot = scipy.sin(theta-self.gamma)
    phiDotdot = thetaDotdot + \
                (thetaDot**2)*sin(phi)-cos(theta-self.gamma)*sin(phi)
    return [thetaDot,thetaDotdot,phiDot,phiDotdot]

self.trajectory = scipy.integrate.odeint(self.dydt, self.GetStateVector(),\
                  timepoints)

def dzdc(self, z, c):
    theta,thetaDot,phi,phiDot,t = z
    y = array([theta, thetaDot, phi, phiDot])
    thetaDot, thetaDotdot, phiDot, phiDotdot = self.dydt(y, t)
    cDot = phiDot - 2.*thetaDot
    return [thetaDot/cDot,thetaDotdot/cDot,phiDot/cDot,phiDotdot/cDot,
            1./cDot]

z = scipy.integrate.odeint(self.dzdc, [y[0],y[1],y[2],y[3],t], 
    scipy.array([self.CollisionCondition(), 0.])

\end{verbatim}

\begin{figure}
\includegraphics[height=2in]{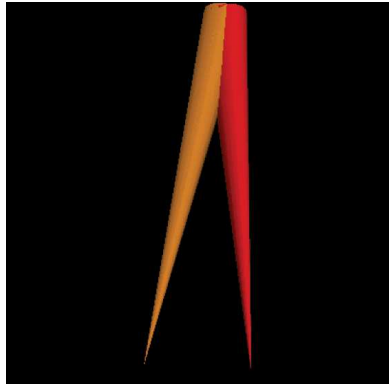}
\caption{\label{WalkerFig}
Snapshot in the perambulation of the Walker.  The model consists of a
pair of coupled pendula (legs) walking down a ramp.  The stance leg (red)
remains fixed with respect to the floor, while the swing leg (orange)
swings forward.  Once the swing leg hits the floor ahead of the stance leg
(heelstrike), the two legs switch roles.  
Real-time animation of the Walker is accomplished using VPython.
}
\end{figure}

The Walker exhibits an interesting period-doubling route to chaos as
the slope of the inclined ramp is increased.  Simple periodic walking
is stable for small ramp angles, but becomes unstable to a period-two
gait at a critical angle.  The period-two orbit bifurcates to a
period-four gait, etc., with increasing angle, culminating in a
chaotic walk.  (The chaos is, however, remarkably subtle, as is also
true in systems like dripping faucets.) A snapshot of the Walker is
shown in Figure 3.  Other modules in our course
enable students to study in considerable more detail these sorts of
period-doubling bifurcations and chaotic dynamics in iterated
one-dimensional maps.

\subsection*{Pattern formation in cardiac dynamics}

Pattern formation is ubiquitous in spatially-extended nonequilibrium
systems.  Many patterns involve regular, periodic phenomena in space
and time, but equally important are localized coherent structures that
break or otherwise interrupt these periodic structures.  Patterns lie
at the root of much activity in living tissues: the regular beating of
the human heart is perhaps our most familiar reminder of the
spatiotemporal rhymicity of biological patterns.  Cardiac tissue is an
excitable medium: rhythmic voltage pulses, initiated by the heart's
pacemaker cells (in the sinoatrial node), spread as a wave through the
rest of the heart inducing the heart muscle to contract, thereby
pumping blood in a coherent fashion.  In some situations, however,
this regular beating can become interrupted by the presence of spiral
waves in the heart's electrical activity (see Figure 4).  These
spiral waves generate voltage pulses on their own, disrupting the
coordinated rhythm of the normal heart, leading to cardiac arrythmia.
A simple model of cardiac dynamics - the two-dimensional
FitzHugh-Nagumo equations\cite{FitzHugh1961, Nagumo1962} - 
is introduced in this course module, which 
we developed in conjunction with Niels Otani.  The FitzHugh-Nagumo model
describes the coupled time evolution of two fields, the transmembrane
potential $V$ and the recovery variable $W$ (given parameters
$\epsilon$, $\gamma$ and $\beta$):
$$
\frac{\partial V}{\partial t} = \nabla^2 V + \frac{1}{\epsilon}
 (V - V^3/3 - W)\ \ \ \ \ \ \ \ \ 
\frac{\partial W}{\partial t} = \epsilon (V - \gamma W + \beta)
$$

Fixed point solutions to the FitzHugh-Nagumo equations are found by 
root-finding, which we accomplish using the \verb+brentq+ function in scipy:

\begin{verbatim}
def FindFixedPoint(gamma, beta):
    f = lambda v, gamma, beta: (v-(v**3)/3.)-((1./gamma)*(v+beta))
    vstar = scipy.optimize.brentq(f, -2., 2., args=(gamma, beta))
    wstar = ((1./gamma)*(vstar+beta))
    return vstar, wstar
\end{verbatim}

We also introduce students to finite-difference techniques for
computing spatial derivatives in the solution of partial differential
equations (PDEs).  Numpy arrays are used to represent the $V$ and $W$
fields of the FitzHugh-Nagumo model, and an important operation is the
computation of the laplacian of the voltage field, $\nabla^2 V(x,y)$.
We introduce the stencil notation for characterizing finite-difference
approximations to $\nabla^2 V$, and use a combination of array
arithmetic and array slicing to compactly and efficiently compute the
derivative on the interior (non-boundary) cells of the simulation
domain.  Students are asked to implement two different approximations
to the laplacian operator (a five-point and nine-point stencil), 
and compare their effects on the detailed form of propagating electrical waves.
The computation of the five-point stencil is shown here:

\begin{figure}
\includegraphics[height=2in]{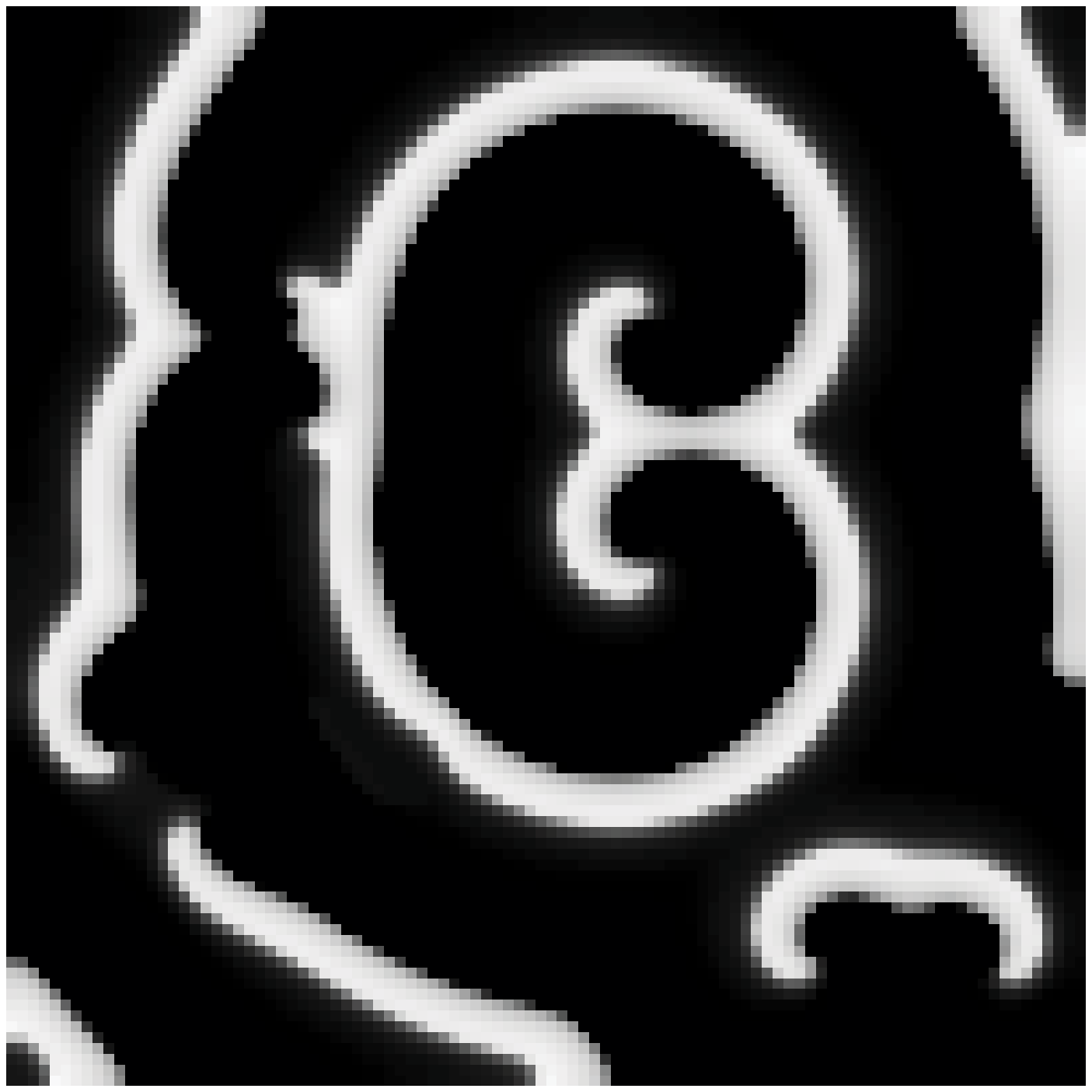}
\includegraphics[height=2in]{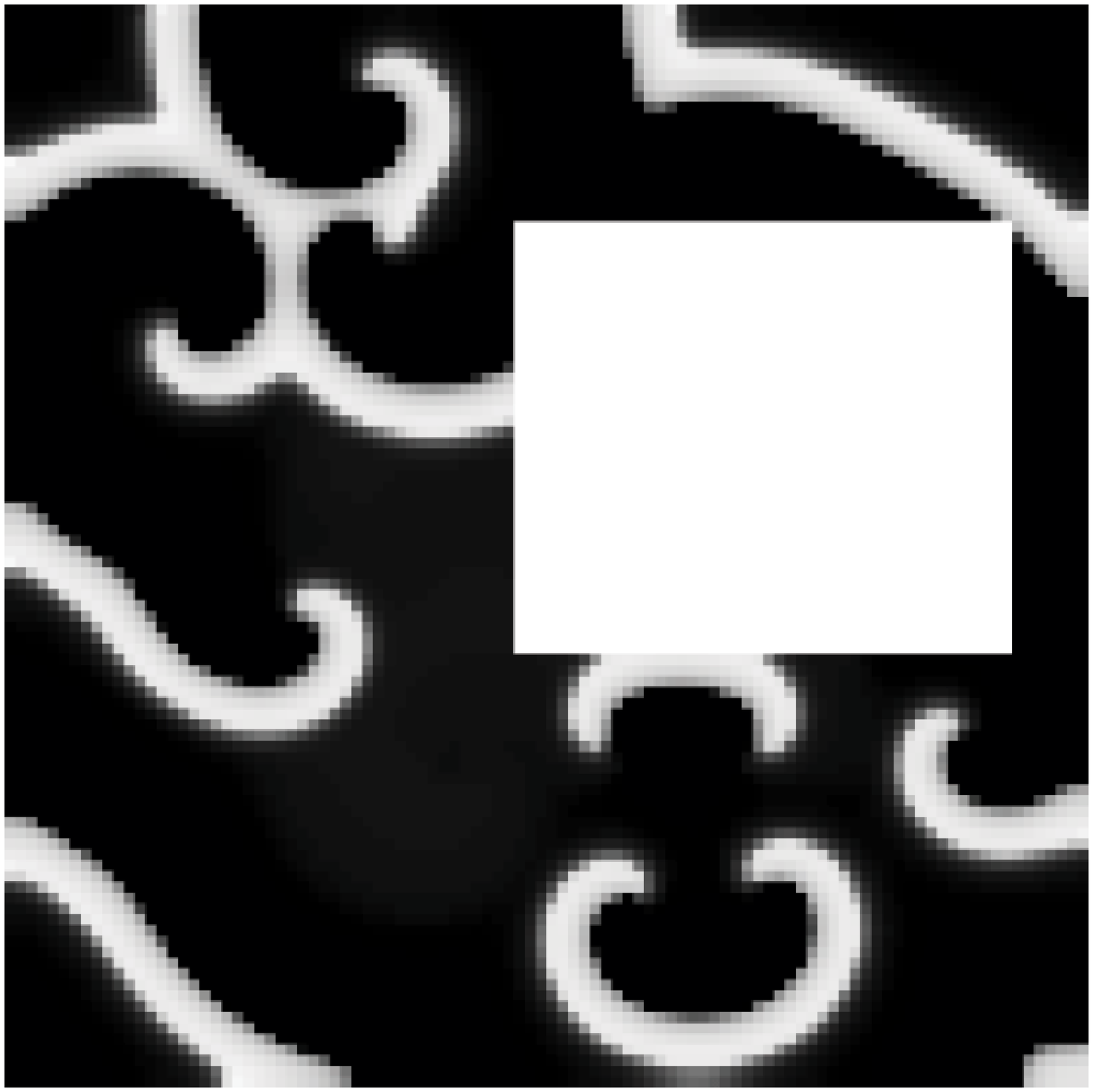}
\caption{\label{CardiacFig}
Snapshots in the time evolution of the FitzHugh-Nagumo model of 
cardiac dynamics.  The transmembrane voltage $V$ is depicted via a 
grayscale map (higher voltages in lighter grays).  Spiral waves in 
the voltage field can lead to cardiac arrythmias by disrupting the 
normal periodic rhythm generated by the sinoatrial node.  (Right) 
Users can administer local voltage pulses (white rectangle) to trigger
spiral wave formation or to shock the arrhythmic heart back to a 
normal beating state.
}
\end{figure}

\begin{verbatim}
def del2_5(a, dx):
    """del2_5(a, dx) returns a finite-difference approximation of the
    laplacian of the array a, with lattice spacing dx, using the five-point
    stencil:
        0   1   0
        1  -4   1
        0   1   0
    """
    del2 = scipy.zeros(a.shape, float)
    del2[1:-1, 1:-1] = (a[1:-1,2:] + a[1:-1,:-2] + \
                        a[2:,1:-1] + a[:-2,1:-1] - 4.*a[1:-1,1:-1])/(dx*dx)
    return del2     
\end{verbatim}

We provide an animation tool that we have written, based on PIL and
Tkinter, that enables students to update the display of the voltage
field V at every time step, and to use the mouse to introduce local
``shocks'' to the system.  (See Figure 4.) These shocks are both
useful in initiating spiral waves and in resetting the global
electrical state of the system as a defribillator might do.  Optional
extensions to the module, developed by our collaborator Otani,
allow for simulations of spontaneous pacemakers, dead regions of
tissue, and more complex heart-chamber geometries, by letting the
various parameters of the model become spatially-varying fields
themselves (again implemented via numpy arrays).

\subsection*{Gene regulation and the Repressilator}

Gene regulation describes a set of processes by which the expression
of genes within a living cell - i.e., their transcription to messenger
RNA and ultimately their translation to protein - is controlled.
While modern genome sequencing has provided great insights into the
constituent parts (genes and proteins) of many organisms, much less is
known about how those parts of turned on and off and mixed and matched
in different contexts: how is that a brain cell and a hair cell, for
example, can derive from the same genomic blueprint but have such
different properties? 

The Repressilator is a relatively simple synthetic gene regulatory
network developed by Michael Elowitz and Stan
Leibler\cite{Elowitz2000}. Its name derives from its use of three
repressor proteins arranged to form a biological oscillator: these
three repressors act in a manner akin to the ``rock-paper-scissors''
game where TetR inhibits $\lambda$ cI, which in turn inhibits LacI,
which in turn inhibits TetR.  A snapshot in the time evolution of the
Repressilator is shown in Figure 5.

\begin{figure}
\includegraphics[height=2in]{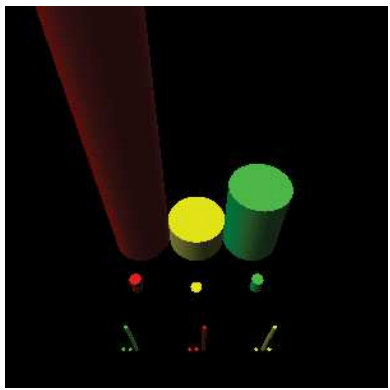}
\caption{\label{RepressilatorFig}
Snapshot in the stochastic time evolution of the Repressilator.  The 
state for this model consists of 15 components: 3 protein concentrations 
(back row), 3 mRNA concentrations (middle row), and 3 sets of promoter-binding
states (front row): each promoter can be either unbound, singly-bound, 
or doubly-bound.  At this instant, TetR (red) concentrations are high, 
leading to suppression of $\lambda$ cI (yellow).  Since $\lambda$ cI is 
low, however, LacI (green) concentrations are allowed to grow.  This will
lead to the eventual suppression of TetR.
}
\end{figure}

One of the important scientific and computational features that we
emphasize in this module are the differences between stochastic and
deterministic representations of chemical reaction networks.  (We
first introduce these concepts in a warmup exercise, Stochastic Cells,
in which students simulate a much simpler biochemical network: one
representation the binding and unbinding of two monomer molecules $M$
to form a single dimer $D$: $M + M \leftrightarrow D$.)  We introduce
students to Petri nets as a graphical notation for encoding such
networks, and then have them, from the underlying Petri net
representation, (a) synthesize differential equations describing the
deterministic time evolution of the system, and (b) implement the
Gillespie algorithm (a form of continuous time Monte Carlo) for
stochastic simulation.\cite{Gillespie1977}  
Gillespie's ``direct method'' involves
choosing a particular reaction and reaction time based on the
instantaneous reaction rates.  For the Repressilator, 
this can be done quite compactly using
array operations within numpy/scipy:

\begin{verbatim}
class StochasticRepressilator (Repressilator):
    # ...
    def Step(self, dtmax):
        self.ComputeReactionRates()
        total_rate = sum(self.rates)
        # get exponentially distributed time
        ran_time = -scipy.log(1.-random.random())/total_rate
        if ran_time > dtmax:
            return dtmax
        # get uniformly drawn rate in interval defined by total_rate
        ran_rate = total_rate*random.random()
        # find interval corresponding to random rate
        reac_index = len(self.rates) - sum(scipy.cumsum(self.rates) > ran_rate)
        reaction = self.reactions[reac_index]
        # execute specified reaction
        for chem, dchem in reaction.stoichiometry.items():
            chem.amount += dchem
        # return time at which reaction takes place
        return ran_time
\end{verbatim}

\subsection*{Other modules}

Our course consists of a number of other modules which we can only
mention in passing here.  As noted, there is a suite of problems
introducing various aspects of chaos and bifurcations in iterated
maps.  There is also a suite of small modules exploring properties of
random walks and extremal statistics.  We have two exercises examining
connections between statistical mechanics and computational
complexity, by probing the nature of phase transitions in NP-complete
problems such as 3SAT.  A random matrix theory module examines the
nature of universality of eigenvalue distributions, and two other
modules explore the thermodynamics of large collective systems (the
Ising model of simple magnets, and the molecular dynamics of large
numbers of atoms).

We continue to look for new problems to add to this collection, and for 
collaborators interested in contributing their scientific and computational 
expertise to this endeavor.  (Please contact us if you have ideas for
interesting modules.)  Our goal is to provide a hands-on introduction
in scientific computing, and it is our hope that this course can help 
serve a number of educational objectives in the part of a larger curriculum 
in computational science and engineering.

\section*{Acknowledgments}

We thank our colleagues who have helped us develop computational
modules and have given us useful feedback: Steve Strogatz, Andy Ruina,
Niels Otani, Bart Selman, Carla Gomes, and John Guckenheimer.  We also
thank all the students who have taken our course and have helped us
work the bugs out of exercises and solutions.  Funding from the NSF
IGERT program (award NSF DGE-0333366) and NSF DMR-0218475 helped
support some initial development of course modules.

\bibliography{CiSE_python_myers_1}

\end{document}